\documentstyle[aps,prl,preprint]{revtex}
\preprint{cond-mat/9402097}
\begin{document}
\title{Systematic $1/S$ study of the 2D Hubbard model at half-filling}
\author{Andrey V. Chubukov and Karen A. Musaelian}
\address{Department of Physics, University of Wisconsin - Madison,\\
1150 University Avenue, Madison, WI 53706\\
and\\
P.L. Kapitza Institute for Physical Problems\\
ul. Kosygina 2, Moscow, Russia 117334}
\date{\today}
\maketitle
\begin{abstract}
The 2D Hubbard model is extended by placing $2S$ orbitals at each lattice
site
and studied in a systematic $1/S$ expansion.
The $1/S$ results for the magnetic susceptibilitity and the
spectra of spin-wave excitations at half filling  are consistent
with the large $S$ calculations for the Heisenberg antiferromagnet.
The $1/S$ corrections to the fermionic spectrum
lift the degeneracy along the edge of the magnetic
Brillouin zone yielding minima at $(\pm\pi/2,\pm\pi/2)$.
Relation to previous papers on the subject is discussed.
\end{abstract}
\pacs{75.10.J, 75.50.E, 05.30}
\narrowtext

\section{Introduction}

 The 2D Hubbard model and the closely related $t-J$ model have
been extensively used recently in the context of
high temperature superconductivity as the simplest models which
capture at least some of the exciting physics of high$-T_c$ superconductors.
At half-filling, the large-$U$ Hubbard model reduces to the nearest-neighbor
Heisenberg antiferromagnet~\cite{t-J}, which perfectly
describes the properties of pure
$La_2CuO_4$~\cite{Birg,CHN,CSY},
while at sufficiently strong doping, it describes a strongly correlated metal
and possibly a superconductor~\cite{Pines,Scalapino}.

One of the crucial issues for lightly doped antiferromagnets is the
role of quantum fluctuations. In $La_{2-x}Sr_{x}CuO_{4}$, they destroy
long-range order in the ground state already at a very small
doping~\cite{Birg}.
There is therefore a need for a  systematic
accounting of quantum fluctuations in the Hubbard
model. For ordered Heisenberg antiferromagnets, a
customary tool for this is the $1/S$ expansion. This
expansion gives rather fast
convergence and first order $1/S$ results are quite
accurate even for $S=1/2$~\cite{oneovers,converge}.

In this paper, we report the results of the large-$S$ studies of quantum
fluctuations in the Hubbard model at half-filling. We view the results as a
basis for future considerations of quantum fluctuations in doped
antiferromagnets.

To perform $1/S$ studies, we need to
extend the Hubbard model to higher spins. A straightforward way to
do this is to place $2S$
equivalent orbitals at each lattice site\cite{haldane,singh}. The hopping term
remains the same as in the conventional Hubbard model, only the hopping
integral $t$ now scales with $S$. At the same time, the $U
n_{\uparrow}n_{\downarrow}$ term
is replaced by
the Hund rule interaction term
\begin{equation}
H_{int} = \frac{U}{2} \sum_{n,\alpha,i}
{\psi}^{\dagger}_{i\alpha n}{\psi}^{\dagger}_{j\beta n}{\psi}_{i\beta n}
{\psi}_{j\alpha n}
\label{eq2}
\end{equation}
Here $\alpha = 1,2$ is the spin index, and $i$ is the orbital index which runs
over $i =1,2....2S$. For $S=1/2$, the summation over spin index is absent and
(\ref{eq2}) reduces to the conventional on-site repulsion term. For $S>1/2$,
the coupling term ensures (at large $U$)
a parallel alignment of the spins of
 electrons at all orbitals, i.e., total spin $S$  at each site,
 which  forms a natural basis for $1/S$
expansion,  and also ensures an
equal number of electrons at each site.  Fluctuations of this number
cost energy of the order of $US \gg t$. The interaction term is simply
related to the total spin per site as $H_{int} = -U \sum_{n} (S^{2}_n +
({{\cal N}_n} -2S)^2) /4$ where ${{\cal N}_n} =
 {\psi}^{\dagger}_{i\alpha n}{\psi}_{i\alpha n}$ is the fermion
density, and ${\bf S}_n = \sum_{i} {\bf S}_{in} =
\frac{1}{2}\psi^{\dagger}_{i\alpha n}{\bf\sigma}_{\alpha
\beta}{\psi}_{i\beta n}$ is the total spin on a given site. The fluctuations of
the square of the total density are irrelevant at large $U$, and we have
checked
explicitly  that one
can equially well use either the eq. (\ref{eq2}) or $-U \sum_{n} S^{2}_n$ for
the interaction term. The former choice is however advantageous for
computational purposes.

In the momentum space, the extended Hubbard Hamiltonian takes the form
\begin{equation}
H = \sum_k \epsilon_k a^{\dagger}_{j\alpha k}a_{j\alpha k} +
\frac{U}{2N}\sum_{k,k',q}a^{\dagger}_{i\alpha,k'}a^{\dagger}_{j\beta,-k'+q}
 a_{i\beta,-k+q}a_{j\alpha,k}
\label{ham2}
\end{equation}
Here $\epsilon_k = -2t(cosk_x+cosk_y)$, $N$ is the total number of lattice
sites, and the momentum summations extend over the first Brillouin zone
 $-\pi < k_x, k_y<\pi$.

The large-$S$ expansion for magnetization and spin susceptibility
of the Hubbard model has been previously
studied in \cite{singh}. Although a complete
agreement with the similar expansion in the Heisenberg model was obtained,
we believe that several relevant contributions were missed,
and the agreement is thus a bit fortuitous.
Below we will point out the discrepancies with Ref.~\cite{singh} explicitly.
The semiclassical expansion results for the magnetization were also
reproduced in ~\cite{chub_fren} by means of a $1/z$ expansion,
 where $z$ is the number of nearest neighbors.

\section{Mean-field theory}

 We now briefly outline the results of
mean-field ($S=\infty$) studies of the Hubbard
model~\cite{tesan,john,bedell,chub_fren,schrief}.
At the mean-field level, one introduces the condensate of $S_{z}(q=Q)$ where
$Q =(\pi,\pi)$,
and uses it to decouple the
interaction term in (\ref{ham2}).
The resulting quadratic Hamiltonian then takes the form
\begin{equation}
H_{\rm {MF}} =  \sum_k \epsilon_ka^{\dagger}_{i\alpha k}a_{i\alpha k} -
 \frac{U<S_z>}{2}
\sum_k a^{\dagger}_{i\alpha,k+Q}\sigma^z_{\alpha\beta}a_{i\beta k},
\end{equation}
where the summations over momentum here and below are limited to the magnetic
zone,
i.e. half the first Brillouin zone.
Note that $<S_z>$ is the exact value of the condensate which includes all
zero-point fluctuations.
This quadratic Hamiltonian can be diagonalized by
means of a Bogolubov transformation
\begin{eqnarray}
c_{i\alpha k} &=& u_k a_{i\alpha k} + v_k \sigma^z_{\alpha\beta}
a_{i\beta,k+Q}\nonumber\\
d_{i \alpha k} &=& v_k a_{i \alpha k} - u_k \sigma^z_{\alpha \beta}
a_{i \beta ,k+Q}
\label{bogolub}
\end{eqnarray}
where $c$ represents conduction, and $d$ --- valence electrons.
 The transformation explicitly reads
\begin{equation}
u_k = \left(\frac{1}{2}\left(1+\frac{\epsilon_k}{E_k}\right)\right)^{1/2};~~~
v_k = \left(\frac{1}{2}\left(1-\frac{\epsilon_k}{E_k}\right)\right)^{1/2},
\end{equation}
where $E_k = (\epsilon_k^2+\Delta^2)^{1/2}$ and $\Delta = \frac{U<S_z>}{2}$.
The diagonalized Hamiltonian then takes the form
\begin{equation}
H = \sum_k E_k(c^{\dagger}_{i \alpha k}c_{i \alpha k}
 - d^{\dagger}_{i \alpha k}d_{i \alpha k}).
\end{equation}
The value of the gap $\Delta$ is obtained from the
self-consistency condition of the diagonalization procedure
\begin{equation}
\frac{1}{U} = \sum_k \frac{2S}{E_k}.
\label{selfcons}
\end{equation}

The bosonic spin-wave excitations appear in the theory
as poles of the total transverse susceptibility
\begin{equation}
{\chi}^{+-}(q,q';\omega) =
i~\int dt~<TS^+_q (t) S^-_{-q'}(0)>
 e^{i \omega t}.
\end{equation}
To zeroth order in $1/S$,  the total static susceptibility is
given by a sum of bubble diagrams in Fig.~\ref{figbubble}:
\begin{eqnarray}
\chi^{+-} = \frac{\chi^{+-}_{0} (q,\omega=0)}{1 - U
{}~\chi^{+-}_{0} (q,\omega=0)}~,
\label{chitot}\\
\chi^{+-}_0 (q,\omega) = \frac{S}{N}\sum_k
\left(1-\frac{\epsilon_k\epsilon_{k+q} -
 \Delta^2}{E_k E_{k+q}}\right)~\frac{2}{E_k+E_{k+q}}
\label{eq3}
\end{eqnarray}
This simple form familiar from paramagnon theories
 exists, however, only for the static susceptibility. The total dynamical
susceptibility is the solution of a $2\times 2$ problem~\cite{schrief} because
the antiferromagnetic ordering doubles the unit cell, and
$\chi^{+-}_0 (q,q';\omega)$
is  nonzero either when $q = q'$ or when $q = q'+Q$ (in the latter case
$\chi^{+-}_{0} (q,q+Q,\omega) \sim \omega$). Also notice the
overall factor of $S$ in $\chi^{+-}_{0}$.
It comes from the summation over the orbital index in each bubble. Because
of this factor, $U \chi^{+-}_{0}$ is $O(1)$, and this makes the
total transverse susceptibility very different from the bare one.

For large $\Delta/t$, one can expand in (\ref{eq3})  and simplify
(\ref{chitot}) to
\begin{equation}
\chi^{+-}(q) = \frac{1}{2J~(1 + \gamma_{q})}
\label{chimf}
\end{equation}
where $\gamma_q = (cosk_x+cosk_y)/2$ and $J=4t^{2}/(2S)^2 U$. As
 $q$ approaches the ordering momentum $Q$, $\gamma_{q} \rightarrow -1$,
 and static susceptibility diverges as it, indeed, should because of the
 Golstone theorem.

\section{ $1/S$ expansion for the static susceptibility}

 We now proceed to the study of the fluctuation corrections to the static
susceptibility. These corrections are related to the residual
interaction between
fermions. In the transverse channel there are two terms --- the direct
interaction of two fermions with opposite spins, and interaction between
fermions of opposite spins mediated by magnetic fluctuations, see
Fig.\ref{figvert}.
The effective Hamiltonian for fermion-magnon coupling was derived
by D. Frenkel and one of us ~\cite{chub_fren}; here
we quote only the result:
\begin{eqnarray}
H_{\rm tr} &=&
\sum_{k,q} (c^{\dagger}_{i\alpha k}c_{i\beta,k+q}e^{\dagger}_{\beta q}
{}~\Phi_{cc} (k,q) +
d^{\dagger}_{i\alpha k} d_{i\beta,k+q}e^{\dagger}_{\beta q}
{}~\Phi_{dd} (k,q) \nonumber \\
&& + c^{\dagger}_{i\alpha k} d_{i\beta,k+q}e^{\dagger}_{\beta q}
{}~\Phi_{cd} (k,q) + d^{\dagger}_{i\alpha k} c_{i\beta,k+q}e^{\dagger}_{\beta
q}
{}~\Phi_{dc} (k,q) +~{\rm H.c.}~)~\delta_{\alpha, -\beta}
\label{tranham}
\end{eqnarray}
Here $e_{\alpha q}$
is the boson annihilation operator with polarization $\alpha$,
and the vertex functions are given by
\begin{eqnarray}
\Phi_{cc,dd} (k,q) &=& \left[\pm (\epsilon_k+
\epsilon_{k+q})\eta_q + (\epsilon_k-\epsilon_{k+q}){\overline\eta}_q
\right]/2S
\nonumber \\
\Phi_{cd,dc} (k,q) &=& U~\left[\left(
1-\frac{(\epsilon_k+\epsilon_{k+q})^2}{8\Delta^2}
\right)\eta_q \mp \left(1-\frac{(\epsilon_k-\epsilon_{k+q})^2}{8\Delta^2}
\right){\overline\eta}_q \right]~,
\label{vertices}
\end{eqnarray}
and $\eta_q$ and $\overline{\eta}_q$ are given by
\begin{equation}
\eta_q = \frac{1}{\sqrt{2}}\left(\frac{1-\gamma_q}{1+\gamma_q}\right)^{1/4},~~
{\overline\eta}_q = \frac{1}{\sqrt{2}}\left(\frac{1+\gamma_q}{1-\gamma_q}
\right)^{1/4}.
\end{equation}
In (\ref{vertices}) we  neglected
all terms of the order $J (t/US)^m \ll J$ with $m\geq 2$ which is
consistent with our intention
to calculate corrections on the scale of magnetic interaction $J$.
Further,  unlike
fermionic energies, the vertex functions
do not have the overall factor of $S$. The second-order self-energy
corrections to the
fermionic propagators then will contain one power of $S$ in the numerator due
to
the summation over the orbitals, and one power of $S$ in the denominator
which comes from  $E_k \propto S$. Total corrections will therefore be
$O(1)$ i.e. will have $1/S$ factor  compared to
 the mean-field inverse  propagators which scale as $O(S)$.

 To calculate the spin susceptibility we will need the
renormalized self-consistency condition, i.e. expression for the order
parameter $<S_z>$ in terms of $\Delta = U <S_z>$. It should ensure the
 cancellation of the leading $O(1)$
term in the denominator  of $\chi^{+-}$. At $S =\infty$, this condition
was given by  eqn. (\ref{selfcons}). For $1/S$ corrections, however, we need
the full expression for $<S_z>$. Substituting (\ref{bogolub})
into the formula for the $z$ component of the spin,
 we obtain after a simple algebra
\begin{equation}
<S_z> = \frac{1}{N}\sum_{i, \alpha, k} \frac{\Delta}{E_k}
(<d^{\dagger}_{i\alpha k} d_{i\alpha k}>
- <c^{\dagger}_{i\alpha k} c_{i\alpha k}>) + \frac{\epsilon_k}{\Delta}
( <c^{\dagger}_{i\alpha k} d_{i\alpha k}> + <d^{\dagger}_{i\alpha k}
c_{i\alpha k}>)
\label{selfc}
\end{equation}
At $S=\infty$, the only nonzero average is
$<d^{\dagger}_{i\alpha k} d_{i\alpha k}> =1$, and using $\Delta = U <S_z>$, we
return to (\ref{selfcons}). However, at finite $S$, one has to calculate pair
averages with the self-energy corrections. These corrections include terms
which transform a conduction fermion into a valence one  and vice versa
after emitting or absorbing a spin wave. Accordingly,
all pair averages, including
$<c^{\dagger}_{i\alpha k} d_{i\alpha k}>$
will be different from zero at finite $S$.
 The density diagrams are presented in Fig~\ref{figdensity}.
 A straightforward calculation using
(\ref{tranham}) and (\ref{vertices})  yields
\begin{equation}
<S_z> = \frac{2S}{N}\sum_k \frac{\Delta}{E_k} -
 \frac{1}{N^2}\sum_{k,p}\frac{4 S^2 \Phi^{2}_{cd}}{(E_k+E_{k+p}+\Omega_p)^2}+
{}~\frac{4J}{U^2}(A-B)~,
\label{selfc2}
\end{equation}
where $\Omega_q = 4JS~\sqrt{1 - \gamma^{2}_q}$ is the bare
spin-wave frequency, and  we defined
\begin{equation}
A = \frac{1}{N}\sum_q\frac{1-\sqrt{1-\gamma_q^2}}{\sqrt{1-\gamma_q^2}},~~
B = \frac{1}{N}\sum_q\frac{\gamma_q^2}{\sqrt{1-\gamma_q^2}}.
\label{defab}
\end{equation}
Retaining only terms of the order of $U$ we arrive at
a well-known result for the Heisenberg model~\cite{oneovers}
\begin{equation}
<S_z> = S\left(1-\frac{1}{SN}\sum_q
\frac{1-\sqrt{1-\gamma_q^2}}{\sqrt{1-\gamma_q^2}}\right)~.
\label{szquant}
\end{equation}

 We are now in a position to compute $1/S$
corrections to the static transverse susceptibility.
Simple considerations show that to the first order in $1/S$,
the RPA approach is still exact, but
one has to include all $1/S$ corrections within a
bubble and also take into account the renormalization of
the relation between $\Delta$ and $U$.
A special care should be given in selecting the diagrams as the
valence-conduction bubble itself has the order of $1/U$. However this leading
contribution gets cancelled in the denominator of (\ref{chitot}), and the
momentum dependence of $\chi^{+-}$ is given by the subleading terms in
$\chi^{+-}_{0}$ which have the
order of $J/U^2$. The cancellation of the $O(U)$ terms should clearly
survive in $1/S$ expansion, and we, therefore, have to keep all $1/S$ diagrams
to the order $J/U^2 S$. Further, one should also consider the effect of
including the
direct longitudinal fermion-fermion interaction. Simple power counting
arguments show  that this
 does not give rise to an extra $1/S$ smallness because each time one
includes an extra interaction, one also has to sum over the
orbitals of intermediate
fermions.
 We found however that in the large-$U$ limit
each inclusion of more than one
triplet ineraction does produce a
smallness, not in $1/S$ but  in $(t/US)^2$,
 because the frequency integration selects only those
combinations of conduction and valence fermions for which the actual
longitudinal interaction becomes small after being dressed by the
Bogolubov coefficients. Altogether, we found 14 relevant diagrams.
They are presented in Fig~\ref{figtranlon} -
\ref{figtwoline}.  The diagrams in Fig.\ref{figtranlon} contain no spin wave
propagator, but two direct fermion-fermion interactions. The diagrams in
Fig.\ref{fignoline} contain self-energy corrections related to the exchange
of spin-waves and no direct fermion-fermion interactions.
The diagrams in Fig.\ref{figoneline} contain one spin-wave propogator
and one direct interaction between fermions with parallel spins. Finally,
the diagrams in Fig.\ref{figtwoline} contain
one spin-wave propagator and two interactions in the triplet channel.
The latter
diagrams were omitted in Ref~\cite{singh} together with several $O(J/U^2 S)$
 terms in other diagrams. We found that altogether,
the terms omitted in~\cite{singh} cancell each
other, but we do not believe that this could be anticipated in advance.
Note that inclusion of self-energy corrections with the direct transverse
interaction
would lead to double-counting, as these corrections have already been taken
into account
in the mean field diagonalization.

Calculation of the diagrams is tedious but straightforward.
We list the expressions in the Appendix and quote here only the final result.
We found that  the modified self-consistency condition ensures an
exact cancellation of
the denominator in (\ref{chitot}) at $q=Q$ in accordance with the Goldstone
theorem.
The functional form of $\chi^{+-}$ is
also unchanged compared to the mean-field result, but quantum fluctuations
give rise to an overall renormalization factor, $Z_{\chi}$. We have
\begin{equation}
\chi^{+-} (q,\omega =0) =
 \frac{Z_{\chi}}{2J(1+\gamma_q)};~~~~Z_{\chi} = 1- \frac{B}{S} =
 1 - \frac{1}{NS}\sum_q\frac{\gamma_q^2}{\sqrt{1-\gamma_q^2}}.
\label{result}
\end{equation}
 As expected, this
result coincides with the known $1/S$ expression for the static susceptibility
of the Heisenberg antiferromagnet~\cite{oneovers}.

Finally, for completeness, we will also obtain the $1/S$ expansion
result for the spin-wave velocity, $c$.
{}From hydrodynamical considerations~\cite{hydro}, we
know that $c^2=\rho_s /\chi_{\perp}~$,
where $\chi_{\perp}$  is the static susceptibility
\begin{equation}
\chi_{\perp} = \frac{1}{2}~ \chi^{+-} (q=0,\omega =0) = \frac{Z_{\chi}}{8J},
\label{chi}
\end{equation}
and $\rho_s$ is the spin stiffness defined by
\begin{equation}
\chi^{xx}_{\rm st}(q \approx Q) = \frac{S_z^2}{\rho_s(q-Q)^2},
\end{equation}
Using the result for $S_z$ and expanding (\ref{result}) near $q=Q$, we find
\begin{equation}
\rho_s = JS^2~Z_{\rho};~~~ Z_{\rho} = 1-\frac{2A}{S}+\frac{B}{S}=
1-\frac{2}{NS}\sum_q
\frac{1-\sqrt{1-\gamma_q^2}}{\sqrt{1-\gamma_q^2}}+\frac{1}{NS}\sum_q
\frac{\gamma_q^2}{\sqrt{1-\gamma_q^2}}
\label{rho}
\end{equation}
Combining (\ref{chi}) and (\ref{rho}), we obtain
\begin{equation}
c^2 = 2J^2~Z^{2}_{c};~~~ Z^{2}_{c} = \frac{Z_{\rho}}{Z_{\chi}} =
1+\frac{2}{S}(B-A)=1+
\frac{2}{NS}\sum_q [1-\sqrt{1-\gamma_q^2}]
\end{equation}
which again coincides with the known spin-wave result.

\section{Self-energy corrections to the fermionic spectrum}
\label{sec_se}

We now turn to the calculation of self-energy corrections to the
single-particle
excitation energy, $E_{k}$. In the mean-field approximation we had
 $E_{k} = \sqrt{\Delta^2 + \epsilon^{2}_k}$. At large $U$, it
can be simplified to
$E_k = \Delta + 4JS (\cos{k_x} + \cos{k_y})^2$.
These forms for $E_k$ imply that
 the minimum of the fermion energy  coincides with the whole
edge of the magnetic Brillouin zone, $\vert k_x\pm k_y \vert = \pi$. This
degeneracy in the position of the minima of $E_k$ was responsible for
the  singular behavior of the Hubbard model at infinitesimally small
doping~\cite{tesan}.
We will show, however, that this degeneracy does not survive perturbative $1/S$
corrections, and the actual minima of the quasiparticle energy are located
only at the particular points $(\pm \pi/2, \pm \pi/2)$ in the Brillouin zone.
This  is consistent with the results of other approaches~
\cite{Vig,Manous,Dag,Trug,Sach}.

Consider for definiteness the corrections to the energy of a valence fermion.
As before, the dominant self-energy corrections at large $U$ are those which
include the exchange of transverse magnetic fluctuations. To first order
in $1/S$,  there are two such diagrams (Fig.\ref{fig_self}). Performing the
frequency integration in these diagrams, we obtain for the Green function of a
valence fermion
\begin{equation}
G^{-1} (k, \omega) = \omega + E_k - \frac{2S}{N} \sum_{q} \left(
\frac{\Phi^{2}_{cd}}{\omega - E_{k+q} -\Omega_q +
i\delta}~+~\frac{\Phi^{2}_{dd}}{\omega + E_{k+q} + \Omega_2 - i\delta} \right)
\label{fgreen}
\end{equation}
We now consider the two self-energy terms separately. The first
term in (\ref{fgreen}) corresponds to a process when valence
fermion transforms into a conduction one after emitting a spin wave. The matrix
element for such process and the energy denominator scale as $U$ and $US$,
correspondingly. As a
result, the first term yields the {\em momentum-independent} renormalizations
of the gap, $\Delta \rightarrow \Delta (1 + A/S)$, and
of the wavefunction residue, $Z = 1 - A/2S$. The latter is in
fact related to the renormalization of the sublattice
magnetization~\cite{chub_fren} $<S_z> = Z -1/2$. This first self-energy
 term also has
momentum-dependent contributions due to the $k$ dependence in $E_{k+q}$, but
they scale as $J$ and, as we will see, are comletely overshadowed by
the momentum-dependent contributions from the second self-energy term.

The second self-energy term in  (\ref{fgreen}) describes a process
which includes only valence fermions. The vertex function for such process
scales as $t/S$ so intuitively one may conclude that the
second term is less relevant than the first one. However, at resonance,
$\omega = -E_k$, and the leading $O(\Delta)$ terms in the denominator get
cancelled. The remaining terms scale as $JS$, so that the total
self-energy correction behaves as $O(\Delta/S)$. More important, however, is
that the $k-$ dependent self-energy terms are not anymore suppressed by a
factor $J/U$ and, therefore, may substantially modify the shape of the
excitation spectrum.

The general structure of the $1/S$ self-energy correction is rather
involved, so below we will restrict calculations to the dispersion
relation along the lines $\vert k_x\pm k_y \vert = \pi$ where the mean-field
energy of valence fermions has a degenerate maximum.
Simplifying eq.(\ref{fgreen}) at $\gamma_k =-1$ and $US/t \gg 1$
 we obtain for the pole of the Green
function
\begin{equation}
E_k^{1/S} = \Delta \left(1 -\frac{1}{S}~+ \frac{\beta(k)}{2S}\right),
\end{equation}
where
\begin{equation}
\beta(k)=\frac{2}{N}\sum_q\frac{1+\sqrt{1-\gamma_q^2}}{2\gamma_{k+q}^2+
\sqrt{1-\gamma_q^2}}~.
\label{beta}
\end{equation}
It is not difficult to see that $\beta(k)$ does contain some
momentum dependence, and, therefore, the excitation energy acquires dispersion
along the edge of the magnetic Brillouin zone. We evaluated the integral
(\ref{beta})
numerically and found that $\beta(k)$ has  four equivalent minima at
 $(\pm\pi/2,\pm\pi/2)$. Near each minimum, one can expand in
${\overline k}_{x,y} = \pm \pi/2 - k_{x,y}$  and obtain the
 dispersion relation typical for the anisotropic 2D Fermi gas:
\begin{equation}
E_k^{1/S}=\overline\Delta+\frac{p_{\perp}^2}{2m_{\perp}} +
\frac{p_{\parallel}^2}{2m_{\parallel}}~,
\label{corr_Ek}
\end{equation}
Here $\overline\Delta$ differs from $\Delta$ due to momentum-independent
self-energy corrections, and we introduced
$p_{\perp} = ({\overline k}_x \pm {\overline k}_y)/\sqrt{2},~
p_{\parallel} = ({\overline k}_x \mp {\overline k}_y)/\sqrt{2}$
(the upper sign is for $(\pi/2,\pi/2)$ and $(-\pi/2,-\pi/2)$ and the
lower sign is for the other two minima). One of the effective masses is
finite already at the mean-field level, and we easily obtain from
$E^{S=\infty}_{k}$:
\begin{equation}
m_{\perp}=\frac{1}{8JS}~(1 + O(1/S))
\end{equation}
On the contrary, $m_{\parallel}$ is infinite at the mean-field level, and
acquires a finite value only due to $1/S$ corrections. The numerical evaluation
of this mass yields
\begin{equation}
m_{\parallel} \approx \frac{2S}{0.084\Delta}.
\end{equation}
For the ratio of the effctive masses
we then obtain to the leading order in $1/S$: $m_{\parallel}/m_{\perp} \approx
188
(JS/U)$. Notice the surprisingly large numerical coefficient.

Care has to be taken in applying the $1/S$ results for the masses to
the physical case of $S=1/2$. We already mentioned that in resonance,
 the diagram with only valence fermions has small energy denominator $O(JS)$
because the leading $\Delta$ term in $E_{k+q}$ gets cancelled by the external
frequency. In this situation,  vertex and self-energy corrections
to the internal fermion line are important as they contain large factors of
$U/J$. A simple analysis similar to that in~\cite{chub_fren} shows
that self-energy and vertex corrections scale as $O(U/JS)$ and $O(U/JS)^2$
respectively. Neglecting these corrections is therefore legitimate only
if $U/JS \ll 1$ and, in a rigorous sense, our $1/S$ expansion for the
fermionic energy is valid only if $S \gg U/J$~\cite{comm}.
It has, however, been argued several times ~\cite{chub_fren,ShSi,KLR} that at
large $U$, the actual
(renormalized) interaction between fermions and spin-waves (which includes
the wavefunction renormalization factor of fermions) retains its functional
form
(at least near magnon momentum $q =(\pi,\pi)$), but
has the same order of magnitude as the
bandwidth. The argument here is that for $U_{eff} \sim JS$, self-energy and
vertex corrections do not generate any new energy scales besides $JS$.
In this situation, the actual expansion parameter for our studies
is $O(1)$, and we can expect that the lowest-order approximation we are using
still yields at least qualitatively correct results.

We conclude this section by a brief comparison with other approaches. Vignale
and Hedayati~\cite{Vig} considered the same perturbative corrections as we did,
but resorted to self-consistent rather than perturbation analysis.
 They required that all corrections to the bare disperson of  quasiparticles
be on the scale of $JS$, and solved self-consistently Dyson equation
for the quasiparticle energy. They found the band minimum at $(\pi/2, \pi/2)$,
which agrees with our finding. We, however, did find the corrections
of the order $O(\Delta/S)$ in perturbative calculations. Boninsegni and
Manousakis~\cite{Manous} used variational Monte-Carlo to compute the
hole dispersion in
$t-J$ model. They also obtained band minima at
$(\pm \pi/2, \pm \pi/2)$. The quantitative comparison with our results is,
however,
difficult because of the relatively small values of $t/J \leq 0.5$ considered
in~\cite{Manous} The minima at $(\pm \pi/2,\pm \pi/2)$ in the $t-J$ model
were also found in the
perturbative calculations in the small $t/J$ limit~\cite{Dag}.
Trugman~\cite{Trug} and Sachdev~\cite{Sach}
performed variational calculations for the Hubbard and $t-J$ models
respectively  and also found the minima
at $(\pm \pi/2, \pm \pi/2)$ at large $t/J$. We compared the bandwidth
along $k_x + k_y = \pi$ and found good agreement between Sachdev's results
and ours. Thus at $t/J =2$, variational and our $1/S$ analysis yield
$\Delta E =  E(\pi,0) -E(\pi/2,\pi/2) \approx 0.2 t$, and $0.24t$
correspondingly.

\section{Conclusions}

 To summarize, in this paper we
considered a large $S$ extension of the Hubbard model
and calculated in $1/S$ expansion
the leading quantum corrections to the static spin
susceptibility, spin-wave velocity and hole dispersion
at half-filling. We found that the mean-field degeneracy of the quasiparticle
spectrum does not survive in $1/S$ calculations, and the actual hole
dispersion has minima at  $(\pm \pi/2, \pm \pi/2)$.
Our results for magnetic parameters agree (as
they should) with the $1/S$ calculations in the Heisenberg model. Indeed, at
half-filling, the use of Hubbard model for magnetic calculations is not the
easiest way to arrive at the final result. However, the method we presented
here has a substantial advantage over other techniques in that
it can be straightforwardly extended to doped
antiferromagnets where quantum fluctuations are
much more relevant than at half
filling. The work along these lines is now in progress.

\section{Acknowledgements}

It is our pleasure to thank D. Frenkel, R. Joynt and S. Sachdev
for useful conversations.
The research was supported in part by the Graduate School at the University
of Wisconsin-Madison and Electric Power Research Institute.

\section{Appendix}

Here we give the expressions  for the diagrams in Fig. \ref{figtranlon} -
\ref{figtwoline}. The two diagrams on Fig. \ref{figtranlon} are obviously
irrelevant to
magnetism, and, accordingly, cancel each other out giving contributions of
$\pm\frac{J}{4U^2S}$.Two of the rest of the diagrams,
 namely those of Fig. \ref{fignoline}a
and \ref{figoneline}a, contribute to order $1/US$,
while the rest contribute to
order $J/U^2 S$. The explicit expressions for the first two diagrams are
\begin{equation}
\chi_{1a} = - \frac{1}{N^2}~\sum_{k,p}~\frac{1}{E_k}~
\left(1-\frac{(\epsilon_k+\epsilon_{k+q})^2}{4\Delta^2}\right)~
\left[\frac{8 S^2 \Phi_{cd}^2}{(E_k+E_{k+p}+\Omega_p)^2}~+~
 \frac{4 S^2 \Phi_{cd}^2}{E_k (E_k+E_{k+p}+\Omega_p)} \right]
\label{A1}
\end{equation}
for the diagram \ref{fignoline}a, and
\begin{equation}
\chi_{2a} = \frac{1}{N^2}~\sum_{k,p}~\frac{4 S^2 \Phi_{cd}^2}
{E^{2}_k (E_k+E_{k+p} +\Omega_p)} ~ - ~ \frac{2J}{U^2 S}~(1+\gamma_q)~A -~
 \frac{2J}{U^2 S}~B
\label{A2}
\end{equation}
for the diagram \ref{figoneline}a.
Here all
the sums are over a half of the Brilloiun
zone, and constants $A$ and $B$ were definied in (\ref{defab}).
Using the self-consistency condition (\ref{selfc2}), the total contribution
from these diagrams can be immediately simplified to
\begin{equation}
\chi_{1a} + \chi_{2a} = \frac{1}{U} ~-~ \sum_{k}~\frac{2S}{E_k}~-~
\frac{2J}{U^2 S}~(2A -B) +~\frac{2J}{U^2 S}~(1 + \gamma_q)~A
\label{A3}
\end{equation}
Care also should be given to the calculation of the bare bubble, because
the relation between $\Delta$ and $U$ contains $1/S$ terms.
{}From ({\ref{eq3}) we have
\begin{equation}
\chi_0 = \chi^{+-}_0 = \sum_{k}~\frac{2S}{E_k}~ - \frac{2J}{U^2}~(1 + \gamma_q)
{}~\left(\frac{US}{\Delta}\right)^3 = \sum_{k}~\frac{2S}{E_k}~ -
\frac{2J}{U^2}~(1 + \gamma_q)\left(1 + \frac{3A}{S}\right)
\label{AA}
\end{equation}
Assembling eqn (\ref{A3}) and (\ref{AA}), we obtain
\begin{equation}
\chi_0 + \chi_1 + \chi_2 = \frac{1}{U}~ \left (1 - \frac{2J}{U}~(1 + \gamma_q)
\right)~-~\frac{2J}{U^2 S}~(2A -B) ~-~~\frac{4J}{U^2 S}~(1 + \gamma_q)~A
\label{A4}
\end{equation}
The expression for other diagrams are as follows
\begin{eqnarray}
\chi_{1b} &=& \frac{2J}{U^2 S}~\frac{1}{N}~\sum_{k} \sqrt{1 -
\gamma^{2}_{k}};~~~\chi_{1c} + \chi_{1d} = \frac{6J}{U^2 S}~(1 +
\gamma_q)~(A-B);~~~\chi_{1e} = \frac{2J}{U^2 S}~(1+\gamma_q)~A \nonumber \\
&& \chi_{2b} = - \frac{2J}{U^2 S}~\gamma_q~(A-B);~~~\chi_{2c} =
\frac{2J}{U^2 S}~\gamma_q~B;~~~\chi_{2d} = \frac{2J}{U^2 S}~(1 +\gamma_q)~B;
\nonumber \\
&& \chi_{2e} = -~\frac{2J}{U^2 S}~\frac{1}{N}~\sum_{k} \sqrt{1 -
\gamma^{2}_{k}};~~~\chi_{2f} = -~\frac{2J}{U^2 S}~(1 +\gamma_q)~A.
\label{A5}
\end{eqnarray}
Finally, the diagrams in Fig~\ref{figtwoline} are
\begin{equation}
\chi_{3a} = \frac{2J}{U^2 S}~A;~~~\chi_{3b} = -~\frac{2J}{U^2 S}~\gamma_q~B.
\label{A6}
\end{equation}
Summing up all contributions and substituting them into eqn. (\ref{chitot}),
we obtain the result quoted in eqn. (\ref{result}).

\begin{figure}
\caption{The RPA series for the total transverse susceptibility
for $S=infty$. The first term represents the
simple bubble, which is the building block of the ladder.}
\label{figbubble}
\end{figure}

\begin{figure}
\caption{The vertex function for the exchange of magnetic fluctuations. This
interaction is obtained from the direct transverse fermion-fermion coupling
after a summation of ladder diagrams.}
\label{figvert}
\end{figure}

\begin{figure}
\caption{The $1/S$ diagrams for the staggered magnetization. The solid line
corresponds to a valence dermion, dashed line corresponds to the conduction
fermion, and the wavy line represents the bosonic spin-wave propagator.}
\label{figdensity}
\end{figure}

\begin{figure}
\caption{The diagram with two direct fermion-fermion interaction lines in both
the transverse and longitudinal channels.}
\label{figtranlon}
\end{figure}

\begin{figure}
\caption{Diagrams for the irreducible susceptibility which do not contain
a fermion-fermion interaction in the longitudinal
channel. A half of the diagrams is presented.
The rest  are equivalent to those presented on
Figs.\protect\ref{fignoline}-\protect\ref{figtwoline} and can be obtained by
 replacing simultaneously all conduction electron propagators by valence
ones and vice versa.}
\label{fignoline}
\end{figure}

\begin{figure}
\caption{Diagrams for the irreducible susceptibility
which contain one direct fermion-fermion interaction line.}
\label{figoneline}
\end{figure}

\begin{figure}
\caption{The same as in Fig.\protect\ref{figoneline}
but with two fermion-fermion interaction lines in the longitudinal channel.}
\label{figtwoline}
\end{figure}

\begin{figure}
\caption{Self-energy diagrams which contribute to the
propagator of valence fermions to order $1/S$.}
\label{fig_self}
\end{figure}

\begin{table}
\caption{Quasiparticle dispersion along $k_x+k_y = \pi$. Energies are
in units of $\Delta$.}
\label{table}
\begin{tabular}{dddddd}
$(\pi/2, \pi/2)$ & $(5\pi/8, 3\pi/8)$ & $(3 \pi/4, \pi/4)$ &
$(7\pi/8, \pi/8)$ & $(\pi,0)$ & \\
\tableline
   0.739 & 0.751 & 0.784 & 0.824 & 0.842 &
\end{tabular}
\end{table}

\begin{references}
\bibitem{comm} For the calculations of the static susceptibility, vertex
corrections of the type discussed in Sec\ref{sec_se} are irrelevant because
the external frequency iz zero (or, more generally, it is $O(JS)$ for all
dynamical magnetic properties).
\end{references}
\end{document}